\documentclass[prl,twocolumn,showpacs,showkeys,superscriptaddress]{revtex4}

\usepackage{graphics}
\usepackage{pslatex}
\usepackage{amsmath}
\usepackage{amssymb}

\include{MyMc} 

\newcommand{\Sec}[1]{}

\begin{document}

\title{%
  Spectroscopy of the fractional vortex eigenfrequency in a long Josephson 0-$\kappa$ junction}

\author{K.~Buckenmaier}
\author{T.~Gaber}
\affiliation{
  Physikalisches Institut -- Experimentalphysik II,
  Universit\"at T\"ubingen,
  Auf der Morgenstelle 14,
  D-72076 T\"ubingen, Germany
}

\author{M. Siegel}
\affiliation{
 Universit\"at Karlsruhe,
 Institut f\"ur Mikro-- und Nanoelektronische Systeme,
 Hertzstr. 16,
 D-76187 Karlsruhe, Germany
}

\author{D.~Koelle}
\author{R.~Kleiner}
\author{E.~Goldobin}
\email{gold@uni-tuebingen.de}
\affiliation{
  Physikalisches Institut -- Experimentalphysik II,
  Universit\"at T\"ubingen,
  Auf der Morgenstelle 14,
  D-72076 T\"ubingen, Germany
}



%

\date{\today}

\begin{abstract}

  Fractional Josephson vortices carry a magnetic flux $\Phi$, which is a fraction of the magnetic flux quantum $\Phi_0\approx 2.07\times10^{-15}\,\mathrm{Wb}$. Their properties are very different from the properties of the usual integer fluxons. In particular, fractional vortices are pinned and have an oscillation eigenfrequency which is expected to be within the Josephson plasma gap. Using microwave spectroscopy, we investigate the dependence of the eigenfrequency of a fractional Josephson vortex on its magnetic flux $\Phi$ and on the bias current. The experimental results are in good agreement with the theoretical predictions.

\end{abstract}

\maketitle

\Sec{Introduction}


In contrast to integer Josephson vortices known for more than 40 years, \emph{fractional Josephson vortices} are relatively new objects. 
Fractional vortices exist in the so-called 0-$\kappa$ long Josephson junctions (LJJs)\cite{Goldobin:2KappaGroundStates,OtherFracVort} consisting of 0-parts, having the usual current-phase relation (CPR) $j_s=j_c\sin(\mu)$, and $\kappa$-parts, having the CPR shifted by $\kappa$, i.e. $j_s=j_c\sin(\mu-\kappa)$. Here $j_s$ and $j_c$ are the supercurrent and the critical current densities and $\mu$ is the Josephson phase. Often $\kappa$ is equal to $\pi$, and one speaks about 0-$\pi$ LJJs, which can be fabricated using various LJJ technologies, e.g. based on d-wave superconductors\cite{VanHarlingen:1995:Review,Kirtley:SF:T-dep,Hilgenkamp:zigzag:SF,Kirtley:2005:AFM-SF} or on a ferromagnetic barrier\cite{Bulaevskii:0-pi-LJJ,DellaRocca:2005:0-pi-SFS:SF,Frolov:2006:SFS-0-pi,Weides:2006:SIFS-0-pi}. 
Using an artificial trick with two tiny current injectors, one can also study 0-$\kappa$ LJJs with arbitrary $\kappa$ (and fractional vortices), which, in addition, can be tuned during experiment\cite{Goldobin:Art-0-pi,Goldobin:2KappaGroundStates}.

While an integer vortex (fluxon) carries one magnetic flux quantum $\Phi_0$, the fractional vortex carries only a \emph{fraction} of the magnetic flux quantum, i.e. a $\kappa$-vortex carries the flux $\Phi=\Phi_0\kappa/(2\pi)$. 

In 0-$\kappa$ LJJs ($\kappa>0$) two types of fractional vortices may exist: a \emph{direct} $-\kappa$ vortex, and a \emph{complementary} $\kappa-2\pi$ vortex\cite{Goldobin:2KappaGroundStates}. The ``smaller'' vortex corresponds to the ground state of the system, while the ``bigger'' one to the excited state. In the case $\kappa=\pi$, both vortices correspond to mirror symmetric semifluxons with a doubly degenerate ground state. Note, that the fluxon ($\kappa=2\pi$) is an excited state of the system, with a constant phase being the ground state. Only using topological protection, e.g., an annular LJJ, one can trap the fluxon reliably.

In contrast to a fluxon, the fractional vortex is pinned at the 0-$\kappa$ boundary. While, being a soliton, the fluxon may freely move along LJJ under the action of various forces (driving, friction, magnetic field gradient), the fractional vortex can only bend/deform, but it always stays in the vicinity of the 0-$\kappa$ boundary. When the forces are released, the fractional vortex recovers its equilibrium shape. If the damping is small, the recovery process is accompanied by decaying oscillations around the equilibrium position with the eigenfrequency, as predicted in Ref.~\onlinecite{Goldobin:2KappaEigenModes}.


It is crucial to know the eigenfrequencies of a fractional vortex. First, any classical device, which uses fractional vortices, should not operate at frequencies in the vicinity of the eigenfrequency (parasitic resonance). Second, an eigenfrequency gives hints about the stability of certain vortex configurations, namely, a low eigenfrequency is a clear sign that the system is close to an instability region, i.e. it may be sensitive to thermal noise, etc. Third, in the quantum domain the eigenfrequency $\omega_0$ defines the attempt frequency of the macroscopic quantum tunneling, while, $\hbar\omega_0$ defines the energy gap between the ground state and the first excited (plasmon) state in the system.

In this letter we report on the first experimental investigation of the eigenfrequency of a fractional Josephson vortex. Using microwave spectroscopy we measure the eigenfrequency as a function of $\kappa$ (or the flux $\Phi$ carried by the vortex) and applied bias current (which deforms the vortex and changes its eigenfrequency). 


\Sec{Theory}

\begin{figure}[!tb]
  \includegraphics{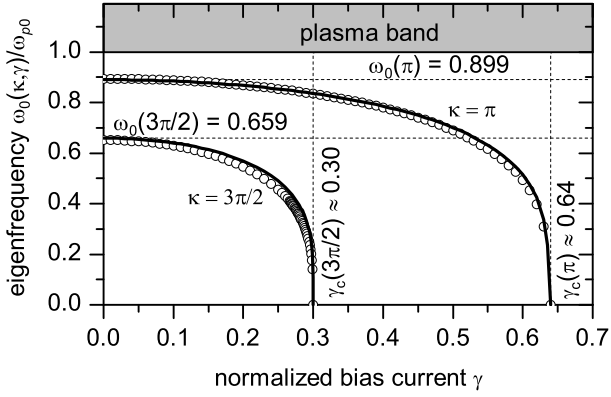}
  \caption{%
    Dependences $\omega_0(\pi,\gamma)$ and $\omega_0(3\pi/2,\gamma)$ obtained by numerically solving the sine-Gordon equation for a 0-$\kappa$ LJJ (symbols) and given by Eq.~(\ref{Eq:EigenFreq(gamma)}) (lines).
  }
  \label{Fig:EigenFreq(gamma)}
\end{figure}

The eigenfrequency $\omega_0$ of a single fractional vortex in an infinite 0-$\kappa$ LJJ at zero normalized bias current $\gamma=I/I_{c0}=0$ and vanishing damping $\alpha=0$ is\cite{Goldobin:2KappaEigenModes}
\begin{equation}
  \omega_0(\kappa,0) = \omega_{p0} \sqrt{
    \frac{1}{2}\cos\frac{\kappa}{4}
    \left(\cos\frac{\kappa}{4} + \sqrt{4 - 3\cos^2\frac{\kappa}{4}}\right)
  }. 
  \label{Eq:EigenFreq0}
\end{equation}
Here $I_{c0}=j_c w L$ is the ``intrinsic'' critical current, $w$ and $L$ are the width and length (circumference) of the LJJ and $j_c$ is the critical current density. In Eq.~(\ref{Eq:EigenFreq0}) the prefactor $\omega_{p0}=\sqrt{\Phi_0 j_c/(2\pi C)}$ is the ``intrinsic'' zero bias plasma frequency related to specific capacitance $C$ of the Josephson barrier and $j_c$. For $\gamma \neq 0$ the analytical expression for $\omega_0(\kappa,\gamma)$ is unknown, but one can approximate it as
\begin{equation}
  \omega_0(\kappa,\gamma) \approx \omega_0(\kappa,0)
  \sqrt[4]{1 - \left(\frac{\gamma}{\gamma_c(\kappa)}\right)^2}
  , \label{Eq:EigenFreq(gamma)}
\end{equation}
where 
\begin{equation}
  \gamma_c(\kappa) 
  = \frac{I_c(\kappa)}{I_{c0}} = \frac{I_c(\kappa)}{j_c w L}
  = \frac{\sin(\kappa/2)}{\kappa/2}
  , \label{Eq:gamma_c(kappa)}
\end{equation}
is the normalized critical current of the junction (depinning current of the fractional vortex) at given $\kappa$\cite{Nappi:2002:ALJJ-Ic(I_inj),Malomed:2004:ALJJ:Ic(Iinj),Goldobin:F-SF}. Approximation (\ref{Eq:EigenFreq(gamma)}) differs from the exact numerical solution by only few percent [except for $\gamma\to\gamma_c(\kappa)$], see Fig.~\ref{Fig:EigenFreq(gamma)}, and follows the same functional dependence as the plasma frequency of a small Josephson junction\cite{Fulton:1974:JJ:ThermEsc,Devoret:1984:JJ:ResAct} 
\begin{equation}
  \omega_p(\gamma) = \omega_{p0}\sqrt[4]{1-\gamma^2}
  , \label{Eq:omega_p(gamma)}
\end{equation}
Thus, the approximation (\ref{Eq:EigenFreq(gamma)}) is exact for $\kappa=0$. These dependences serve as a guide for planning and performing the experiment. 

\Sec{Experiment}

\begin{figure}[!b]
  \includegraphics{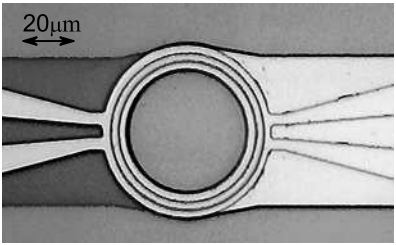}
  \caption{%
    Optical image of the investigated sample: ALJJ with two pairs of injectors.
  }
  \label{Fig:sample}
\end{figure}
%

For the experiments we used tunnel Nb-AlO$_x$-Nb annular LJJ (ALJJ) equipped with two pairs of current injectors as shown in Fig.~\ref{Fig:sample}.   The tunnel ALJJ has low damping and allows us to perform resonant excitation of the fractional vortices and perform spectroscopy. The annular geometry, where the total topological charge is fixed, prevents flipping of a direct vortex to a complementary one with emission of a fluxon. Even if this happens, upon reset to the zero voltage state, the fluxon will be reabsorbed, turning the vortex into the initial state. Using the injectors we can change the value of $\kappa$ during experiment and investigate $\omega_0(\kappa)$. 

The ALJJ has a mean radius $R=30\units{\mu m}$, the width of the injectors and the distance between them are $5\units{\mu m}$. The Josephson penetration depth $\lambda_J\approx 43\units{\mu m}$ was estimated taking into account the idle region\cite{Monaco:1995:IdleReg:Dyn}. The normalized length of the ALJJ, thus, is $\ell\approx 4.35$.

\begin{figure}[!tb]
  \includegraphics{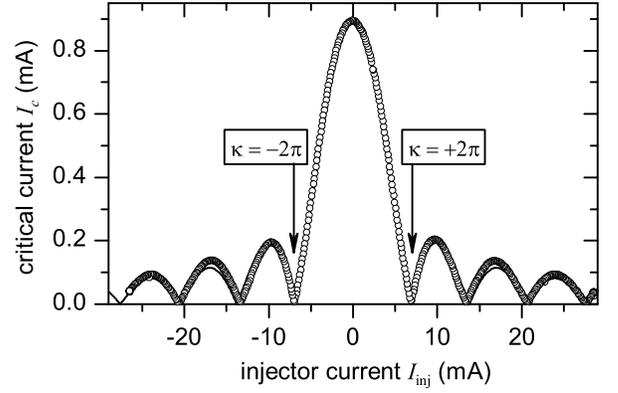}
  \caption{%
    The dependence $I_c(I_\mathrm{inj})$ measured at $T\approx4.2\units{K}$ (symbols) and corresponding theoretical curve (continuous line).
  }
  \label{Fig:Ic(Iinj)}
\end{figure}

To calibrate injectors we have measured the critical current $I_c$ as a function of $I_\mathrm{inj} \propto \kappa$ for the left injector pair in Fig.~\ref{Fig:sample}. This $I_c(I_\mathrm{inj})$ is presented in Fig.~\ref{Fig:Ic(Iinj)} and looks like a Fraunhofer pattern in accordance with the theory\cite{Nappi:2002:ALJJ-Ic(I_inj),Malomed:2004:ALJJ:Ic(Iinj),Goldobin:F-SF}. The right injector pair shows equally good $I_c(I_\mathrm{inj})$, but was not used in this experiment. Knowing $I_\mathrm{inj}$, the value of $\kappa$ can be calculated as $\kappa = 2\pi I_\mathrm{inj}/I_\mathrm{inj}^\mathrm{min}$, where $I_\mathrm{inj}^\mathrm{min} \approx 6.92\units{mA}$ is the injector current corresponding to the first minimum of $I_c(I_\mathrm{inj})$\cite{Goldobin:Art-0-pi,Gaber:Art2}.

\begin{figure}[!t] 
  \begin{center}
    \includegraphics*{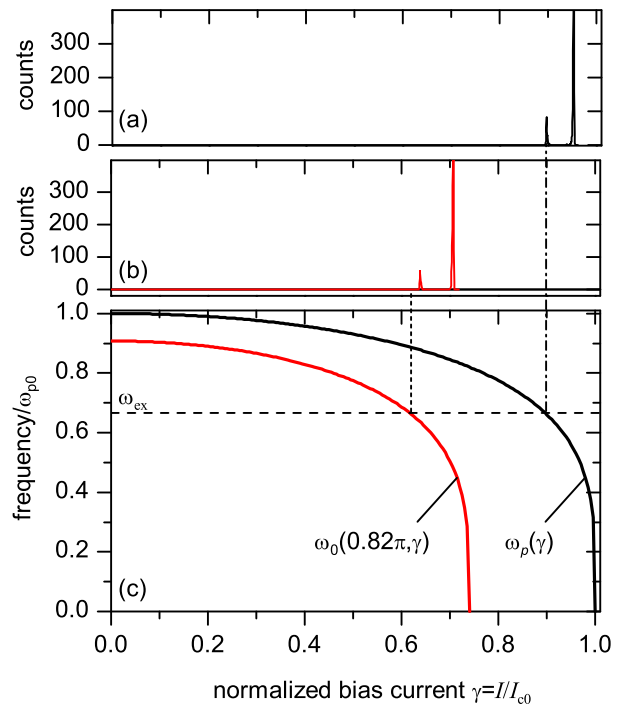}
  \end{center}
  \caption{(Color online)
    Principle of the measurements. The escape histogram measured at $\omega_\mathrm{ex}/2\pi=29\units{GHz}$ for (a) $\kappa=0$ and (b) $\kappa=0.82\pi$ shows two peaks. (c) shows the numerically simulated dependences $\omega_0(0.82\pi,\gamma)$ (\ref{Eq:EigenFreq(gamma)}) and $\omega_p(\gamma)$  (\ref{Eq:omega_p(gamma)}). Applied frequency $\omega_\mathrm{ex}$ is shown by horizontal line.
  } 
  \label{Fig:EigenFreqExp}
\end{figure}

The measurements of the eigenfrequency were performed using resonant excitation of the fractional vortex by microwaves as follows. We apply a microwave radiation of fixed frequency $\omega_\mathrm{ex}$ and power $P$ to our 0-$\kappa$ ALJJ with a $\kappa$-vortex. Then we ramp the bias current $I$ from zero up to above $I_c$ and observe at which current $I_1$ our ALJJ switches to non-zero voltage. Since, according to Eq.~(\ref{Eq:EigenFreq(gamma)}), the eigenfrequency decreases with the bias current, by ramping $I$ one may reach the resonant condition $\omega_0(\kappa,\gamma)=\omega_\mathrm{ex}$ at some $\gamma_1=I_1/I_{c0}$. The fractional vortex will be resonantly excited and will switch the ALJJ to the voltage state. In fact, we have been ramping the bias current many times, and have measured the escape histogram, which represents the probability of switching to the voltage state as a function of bias current $\gamma$, see Fig.~\ref{Fig:EigenFreqExp}a,b. These histograms typically have two maxima. The first maximum (at smaller $I$) corresponds to the resonant excitation of the phase $\mu(x)$ (in our case fractional vortex or phase-particle from the tilted washboard potential) by microwaves. The second maxima (at larger $I$) corresponds to the thermal escape of the Josephson phase at $I$ approaching $I_c(\kappa)$. 

Since the accuracy of our measurements is defined by the width of the peaks in the escape histograms, special measures were taken to suppress electronic noise down to the level that the width is due to the bath temperature. We have used a copper box directly around our junction to shield electromagnetic radiation and a cryoperm shield against static magnetic fields. In addition, the measurements were made in a cryostat surrounded by a three layer $\mu$-metal shield. The whole setup was placed in an electromagnetically shielded room. Bias lines contained three-stage low pass filters with a $3\units{dB}$ level cut-off frequency at $30\units{kHz}$ (by design): two cold filters are in the vicinity of the sample and a warm one at the top of the dip-stick. 
The measurement technique and setup are similar to those described in Ref.~\onlinecite{Wallraff:2003:LJJ:ThermEsc}. The bias current was supplied by a ramp generator, which starts the ramp at a small negative value of current and sends out a digital pulse when the ramp crosses zero level. When the ALJJ goes to the non-zero voltage state, the voltage detector (embedded in the ramp generator) sends out another pulse. By measuring the time between these two pulses and knowing the ramp rate, we  calculate the switching current with high accuracy.
The microwave current is induced in the bias leads of the ALJJ by means of an antenna placed just above the chip inside a copper box and connected to the rf generator. A $10\units{dB}$ attenuator is placed right in front of the copper box to protect the sample from external noise carried through the semirigid microwave cable. Furthermore, the electronics inside the shielded room is battery powered and optically decoupled from the outside\cite{Wallraff:2003:LJJ:ThermEsc}. With this setup we can obtain a histogram of $10^4$ switching events in about $3\units{min}$. The current resolution of this setup is about $30\units{nA}$ which is limited by the noise (jitter) of the current source.

Before measuring the eigenfrequency of a vortex and comparing it with the theory, we have to measure the plasma frequency $\omega_{p0}$ with reasonable accuracy. To do this we excite our system by microwaves and make escape measurements at $I_\mathrm{inj}=0$. In this case the ground state of the system corresponds to a uniform phase $\mu(x)=\arcsin(\gamma)$ and the lowest eigenfrequency is the plasma frequency $\omega_p(\gamma)$, Eq.~(\ref{Eq:omega_p(gamma)}), corresponding to spatially uniform oscillations (like in a point-like junction). 
For a given microwave frequency $\omega_\mathrm{ex}$ the escape histogram has two peaks shown in Fig.~\ref{Fig:EigenFreqExp}a. The first peak at the intersection of $\omega_p(\gamma)$ and $\omega_\mathrm{ex}$ (Fig.~\ref{Fig:EigenFreqExp}c) corresponds to a resonant excitation of the phase from the potential well in a tilted washboard potential\cite{Devoret:1984:JJ:ResAct}. The second peak corresponds to the thermal escape of the phase from the potential well when the well becomes very shallow\cite{Fulton:1974:JJ:ThermEsc,Devoret:1984:JJ:ResAct}. Its form depends on the bias current ramp rate, on the bath temperature $T$ and on the electronic noise\cite{Silvestrini:1988:UnderdampedJJ:IcDistr}. Higher $T$ leads to a broader peak shifted towards lower currents, which obscures the plasma frequency peak especially for low $\omega_\mathrm{ex}$ (high $\gamma$). The electronic noise has, to the first order, the same effect as the increase of $T$. Therefore, our setup was optimized as described above to keep the broadening caused by the electronic noise below the one caused by $T$.

The power $P$ of the applied microwaves was kept as low as possible, but such that the first peak in the histogram is still visible. By increasing $P$ the height of the first peak increases, but its position shifts due to the non-linearity of the resonance. Accordingly, the height of the second peak decreases and it also shifts to lower $\gamma$. Measurements turned out to be easier at lower power and large $\gamma$ when the particle is in the shallow well and both peaks are close. On the other hand, to resonantly excite the particle from the deep well (small $\gamma$), high power, which leads to non-linear effects, is needed. Therefore the measurement results for high $\omega_\mathrm{ex}$ (low $\gamma$) are not so accurate as for low $\omega_\mathrm{ex}$ (high $\gamma$).

Escape histograms were measured for different $\omega_\mathrm{ex}$ and the position of the first peak $I_1(\omega_\mathrm{ex})$ was, after inverting from $I_1(\omega_\mathrm{ex})=I_1(\omega_p)$ to $\omega_p(I_1)=\omega_p(\gamma)$, fitted using the theoretical dependence (\ref{Eq:omega_p(gamma)}). The fitting gives $\omega_{p0}/2\pi=42.73\units{GHz}$ and a noise free $I_{c0}=961\units{\mu A}$. Note that $I_{c0}$ represents the value of $j_c$ most accurately. The critical current $I_c = I_s^\mathrm{max} = 936\units{\mu A}$ measured from the thermal escape peak without microwaves represents the maximum supercurrent and has a somewhat lower value because of the slightly non-uniform current distribution in our ALJJ geometry. The critical current $I_c^\mathrm{IVC}=895\units{\mu A}$ measured from the $I$--$V$ characteristic or from $I_c(I_\mathrm{inj})$ at $I_\mathrm{inj}=0$ (Fig.~\ref{Fig:Ic(Iinj)}) is even lower.

An independent numerical study of a microwave driven point-like Josephson junction predicts that the position of the first peak is somewhat below the plasma resonance\cite{Gronbech-Jensen:2004:JJ+uwave:ThermEsc} calculated using Eq.~(\ref{Eq:omega_p(gamma)}). For LJJ the situation may be even more complicated as the effective 1D potential may depend on various parameters, e.g., externally applied magnetic field\cite{Castellano:1996:LJJ:ThermEsc} or trapped flux. For ALJJ there are no theoretical calculations or numerical simulations so far. In this work we assume that $\omega_p(\gamma)$ is well described by Eq.~(\ref{Eq:omega_p(gamma)}) even for ALJJ.

\begin{figure}[!tb] 
   \begin{center}
     \includegraphics{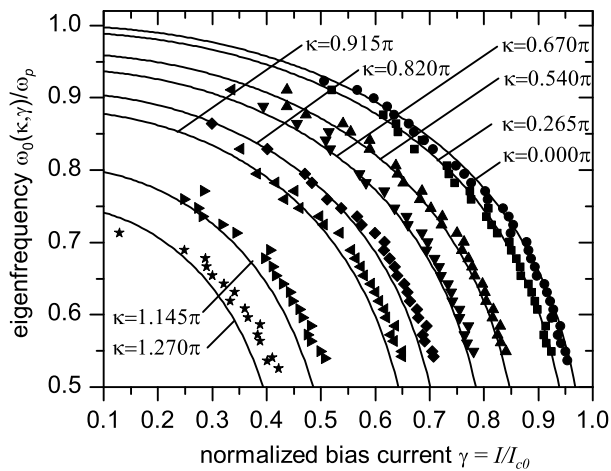}
   \end{center}
   \caption{%
     Comparison between the theoretical curves with measurements of the eigenfrequency.
   } 
   \label{Fig:EigenFreq:EvsT}
\end{figure}

To measure the eigenfrequency of a $\kappa$-vortex we repeated escape measurements for $\kappa\propto I_\mathrm{inj}>0$. Again, the histogram contains two peaks, as shown in Fig.~\ref{Fig:EigenFreqExp}b. The first peak corresponds to a resonant escape of a fractional vortex at $\omega_0(\kappa,\gamma)=\omega_\mathrm{ex}$ (Fig.~\ref{Fig:EigenFreqExp}c). The second peak corresponds to the thermally stimulated switching of the junction at the pre-critical current. Note, that the critical current of our ALJJ depends on $\kappa$ as described by Eq.~(\ref{Eq:gamma_c(kappa)}) and shown in Fig.~\ref{Fig:Ic(Iinj)}. We have measured many escape histograms for a range of $\omega_\mathrm{ex}$ and $I_\mathrm{inj}$ (i.e. $\kappa$), and were able to plot $\omega_0(\kappa,I)$. Knowing $\omega_{p0}$ and $I_{c0}$ from the plasma frequency measurements we have plotted the dependence $\omega_0(\kappa,\gamma)$ in normalized units in Fig.~\ref{Fig:EigenFreq:EvsT}. The theoretical curves $\omega_0(\gamma)$ at different $\kappa$ are shown for comparison. To simulate them, we, first,  numerically solved the sine-Gordon equation for an ALJJ with injectors of finite width to find the static state of the system. Then, to find the eigenfrequency, we have solved the associated eigenvalue problem numerically and,  among all eigenfrequencies, selected the lowest positive one (imaginary part of eigenvalue). 

Fig.~\ref{Fig:EigenFreq:EvsT} shows that the experimental results are in good agreement with the theoretical predictions. Comparing the simulation results for finite injector sizes with the ones for ideal (point-like) injectors, we saw that the eigenfrequency in Fig.~\ref{Fig:EigenFreq:EvsT} shifts towards larger $\gamma$ (or $\omega_0$ increases for fixed $\gamma$) as injectors get larger. Although we have included the geometrical dimensions of our injectors in simulations presented in Fig.~\ref{Fig:EigenFreq:EvsT}, one can see that the experimental points are still somewhat shifted towards larger $\gamma$. This indicates that the effective injector size is larger than its geometric size.

\Sec{Conclusions}

In summary, we have performed spectroscopy of the fractional vortex eigenfrequency $\omega_0(\kappa,\gamma)$ in an annular long Josephson tunnel junction. The results are in agreement with the model. For low values of $\gamma$ the accuracy of our measurements is limited by non-linear effects. For large $\gamma$ (low $\omega_\mathrm{ex}$) the measurements are limited by the thermal width of the peaks in the escape histogram. To study $\omega_0(\kappa,\gamma)$ for $\gamma\to1$ we are going to perform the same measurements at $T<4.2\units{K}$. Another interesting topic is the spectroscopy of a molecule made of two coupled $\kappa$-vortices, in which eigenfrequency splitting is expected. 

\begin{acknowledgments}
  We are grateful to the group of Prof. A. Ustinov (especially to A. Kemp) from University of Erlangen for sharing their experience and numerous suggestions for improving the escape measurements setup. 
  This work is supported by the Deutsche Forschungsgemeinschaft (projects GO-1106/1 and SFB/TR-21) and by the Elitef\"orderprogramm of the Landesstiftung Baden-W\"urttemberg.
\end{acknowledgments}

\bibliographystyle{apsprl}
\bibliography{SF,pi,JJ,LJJ} 


\end{document}